\documentclass[pre,aps,twocolumn,showpacs,amsmath,amssymb]{revtex4}

\usepackage{graphicx}
\usepackage{dcolumn}

\begin{document}
\def \b{$\beta$ }
\def \d{$D$ }
\def \p{$\rho$ }
\def \H{$H$ }
\def \ps{$\|X(w)|^{2}$ }
\def \~{$\approx$ }


\title{Fractal analysis on internet traffic time series}

\author{K. B. Chong$^{a,b, *}$, K. Y. Choo$^{c,b}$}
\address{$^{a}$Department of Physics, National University of Singapore, 10 Kent Ridge Road, Singapore 119260 \\ $^{b}$School of Physics, Universiti Kebangsaan Malaysia, 43600 Bangi, Selangor, Malaysia\\ $^{c}$Faculty of Engingeering, Multimedia University, 63100 Cyberjaya, Selangor, Malaysia}
\email{scip1295@nus.edu.sg}

\begin{abstract}
Fractal behavior and long-range dependence have been observed in
tele-traffic measurement and characterization. In this paper we
show results of application of the fractal analysis to internet
traffic via various methods.  Our result demonstrate that the
internet traffic exhibits self-similarity,  and giving the
spectral exponent (\b : 1 $<$ \b $<$ 2).  Our analysis showed that
Holder exponent (\H :  0 $<$ \H $<$ 0.5) ,  fractal dimensions (\d
: 1 $<$ \d $<$ 2) and the correlation coefficients are (\p : -1/2
$<$ \p $<$ 0). Time-scale analysis show to be an effective way to
characterize the local irregularity.  Based on the result of this
study, these two Internet time series exhibit fractal
characteristic with long-range dependence.
\end{abstract}


\maketitle

\section{\label{I} INTRODUCTION}
Fractal behavior and long-range dependence have been observed in
many phenomena, basically in the field of fluctuations in physical
systems such as diffusion\cite {1,4,6,13,15,17}, financial time
series\cite {7}, tele-traffic\cite {13,16,17,18,19} and time
series of heart rate dynamic\cite {10,17} and human gait\cite
{10}. In this paper, we characterize the dynamics of internet
traffic time series.  We applied fractal analysis into the
internet traffic time series via various methods, such as
power-spectral analysis(PSA), detrended fluctuation analysis(DFA)
and time-scale analysis(TSA).\\ Data to be analyzed are document
sizes which are transferred through Library Of Congress (LOC) WWW
server.  Two types of Internet traffic, namely LOC(request) and
LOC(send) are examined in this paper. LOC(request) is the time
series of document sizes which transferred into the server where
LOC(send) is the time series of document sizes which transferred
out from the sever.  These internet traffic time series are plays
an important role in determining the degree of smooth ascessing
via a particular server.\\ The presence of "burstiness" across an
extremely wide range of time scale in both the time series showed
that both of these internet traffic time series are different from
conventinoal model for telephone traffic, i.e.. pure Poisson or
Poisson-related formal model for packet traffic\cite
{14,16,18,19}.

\section{\label{II} SOME PROPERTIES OF FRACTAL}
Fractal characterizes the object or process by using a fractional
geometry or simplify Fractal geometry, \d.  A fractal object can
be characterize by a dimension between two integers, i.e.. \d =
1.5 . Fractal have following two important properties:-\\
  (a) Self-similarity or self-affine.  A Fractal object similar with
other part even for different scales.  This property namely
scale-invariance which fractal object will similar in all possible
scales.  Self-similar exist when the object show similarity under
isotropic scaling, meanwhile self-affine exist when object show
similarity under an inisotropics scaling.\\
  (b) Self-similar hierarchy structure under magnification.  A
fractal object consist complex inner structure and show similar
geometry even under different magnatification scale\cite {15}. Due
to the scale invariance, a power-law behavior exist in between two
parameters in a  fractal phenomenon , like
\begin{eqnarray}
f(x)\propto x^{c}, \label{eq:1}
\end{eqnarray}
where $f(x)$ is a function of a study object and c is a constant.
From the example given by \cite{20}, one can estimate the fractal
dimension through this power-law behavior.\\
  Standard definition of fractional Brownian motion are introduced
by Mandelbrot and Van Ness\cite{17} and given by:
\begin{eqnarray}
B_{H}(t)&=&\frac{1}{\tau(H+\frac{1}{2})}\bigg[\int^{0}_{-\infty}\{(t-s)^{H-\frac{1}{2}}-(-s)\}dB(s)+\nonumber
\\&& \int^{t}_{0}(t-s)^{H-\frac{1}{2}}dB(s)\bigg], \label{eq:2}
\end{eqnarray}
with Holder exponent, 0 $<$ \H $<$ 1 .  Fractional Brownian motion
consist the following properties:-
\begin{eqnarray}
E[B_{H}(t)]=0, \label{eq:3}
\end{eqnarray}
\begin{eqnarray}
E[B_{H}(t)^{2}]&\sim&t^{2H}, \label{eq:4}
\end{eqnarray}
\begin{eqnarray}
E[B_{H}(t)B_{H}(s)]&=&\frac{1}{2}\left[|t|^{2H}+|s|^{2H}-|t-s|^{2H}\right],\label{eq:5}
\end{eqnarray}
From Eq.(4), the correlation between increment for ${B_{H}(t)}$
can be written in equation form.  For fractal processes, \p can
defined as,
\begin{eqnarray}
\rho&=&\left\langle\frac{-B_{H}(-t)B_{H}(t)}{B_{H}(t)^{2}}\right\rangle,\label{eq:6}
\end{eqnarray}
\begin{eqnarray}
\therefore \rho&=&2^{2H-1}-1,\label{eq:7}
\end{eqnarray}
Where ${B_{H}(t=t_{0}) = 0}$, ${B_{H}(t= -t)= B_{H}(-t)}$, and
${B_{H}(t) = B_{H}(t)}$. If $y(t)$ is a fractal process with
Holder exponent \H , and then for arbitrary process with
\begin{eqnarray}
y(ct)&\triangleq&c^{H}y(t),\label{eq:8}
\end{eqnarray}
also is a fractal process with same statistical distribution,
where $c$  is a constant and  ${c > 0}$. The fractal dimension,
are given by
\begin{eqnarray}
D&=&2-H,\label{eq:9}
\end{eqnarray}
and table I give the relationships for \H , \d, correlation and
the process behavior.
\begin{table}
\caption{\label{Tab:1} Different value $H$ and $D$ and their
associated process}
\begin{ruledtabular}
\begin{tabular}{|c|c|c|c|}
  \H & \d & correlation & process behavior \\ \hline
  $>$0.5 & $<$1.5 & positive & persistence \\
  $=$0.5 & $=$1.5 & zero     & Brownian motion \\
  $<$0.5 & $>$1.5 & negative & anti persistence \\ \hline
\end{tabular}
\end{ruledtabular}
\end{table}

\section{\label{III} POWER-SPECTRAL ANALYSIS (PSA)}
    A time series can be described in time domain as $x(t)$ and also
in frequency domain in term of Fourier transform as $X(\omega)$
where $\omega$ is frequency.  The autocorrelation function of a
non-stationary time series is given as,
\begin{eqnarray}
R_{xx}(t+\tau)&=&\int^{+\infty}_{-\infty}E[x(t)x(t+\tau)]dt,\label{eq:10}
\end{eqnarray}
and the Fourier transform of this autocorrelation function same
with $|X(\omega)|^{2}$, therefore the power-spectral density of a
time series can be written as,
\begin{eqnarray}
S(\omega)&\triangleq&|X(\omega)|^{2},\label{eq:11}
\end{eqnarray}
also Wiener-Kintchine theorem expresses the relationship between
the Fourier transform of the autocorrelation function and
power-spectral density of a time series, as
\begin{eqnarray}
R_{xx}&\longleftrightarrow&S(\omega),\label{eq:12}
\end{eqnarray}
    The power-spectral function provide an important parameter which
characterize the persistency in time series.  For a self-affine
time series, the power-spectral obey the frequency based power-law
behavior, and given by
\begin{eqnarray}
S_{m}(\omega)&\sim&\omega_{m}^{-\beta},
m=1,2,...,\frac{N}{2},\label{eq:13}
\end{eqnarray}
where $\omega_{m}=\frac{m}{N}$; $N$ is length of time series and
spectral exponent , \b characterizes the persistency.  The
relationship between the \b, \H and \d is given by
\begin{eqnarray}
\beta&=&2H+1=5-2D, \label{eq:14}
\end{eqnarray}
    Least-square best fit line are applied in the power-spectral to
get the value of \b.  PSA only provide the value of global Holder
exponent , \H since Fourier transform using harmonic function. PSA
was a conventional methods in fractal analysis since it convenient
to estimate the value of \H \cite{8}.

\subsection*{RESULT(A)}
The power-spectral exponent \b, Holder exponent \H, fractal
dimension \d and correlation coefficient \p of the LOC(request)
and LOC(send) estimated with PSA method, and tabulated in table
II. And also Fig. 1 show the power-spectral for the LOC(request)
and LOC(send) time series.  PSA showed that these LOC(request) and
LOC(send) exhibit fractal characteristic with long-range
dependence.
\begin{table}
\caption{\label{Tab:2} ${\beta, H, \rho}$ and $D$ for LOC(request)
and LOC(send).}
\begin{ruledtabular}
\begin{tabular}{|c|c|c|c|c|}
  Time series & \b & \H & \p & \d \\ \hline
  LOC(request) & 1.59$\pm$0.01 & 0.30$\pm$0.01 & -0.24$\pm$0.01 & 1.70$\pm$0.01 \\
  LOC(send) & 1.61$\pm$0.01 & 0.31$\pm$0.01 & -0.23$\pm$0.01 & 1.69$\pm$0.01 \\ \hline
\end{tabular}
\end{ruledtabular}
\end{table}

\section{\label{IV} DETRENDED FLUCTUATION ANALYSIS (DFA)}
    Detrended fluctuation analysis (DFA) has been widely used to
determine mono-fractal scaling properties and long-range
dependence in noisy, nonstationary time series.  DFA is used to
estimate the root-mean-square fluctuation of an integrated and
detrented time series (a modified root-mean-square analysis of
random walk), and had the capability of detection of long range
dependence. The mathematical form of the integrated time series
$Y(i)$ is denoted as\cite {5},
\begin{eqnarray}
Y(i)&\equiv&\sum^{i}_{k=1}[x_{k}-<x>]; i=1,....,N, \label{eq:15}
\end{eqnarray}
where ${x_{k}}$ is $k$-sequence of the time series, and $<x>$ is
the average of the time series of length $N$.\\
    Next, $Y(i)$ is deviated into ${N_{s}\equiv int\frac{N}{s}}$ non-overlapping segments
of equal length s as shown in Fig. 2. Since, the length of the
time series is often not a multiple of time scale $s$, a short
part at the end of the integrated time series may remain. To
overcome this problem, the same procedure is repeated starting
from the opposite end, and the remain part of the time series is
analyzed too. Therefore, the total segments are 2${N_{s}}$. After
the integrated time series is deviated into ${N_{s}}$ segments,
which each segment has the same equal length $s$, a least-square
best fit line is fitted onto the time series to obtain the local
trend in that particular segment as shown in Fig.2.\\
    The detrending of the time series is done by the subtraction of
the least-square best fit line from the integrated time series,
and variance of each segment is calculated by
\begin{eqnarray}
F^{2}(s,\nu)&\equiv&\frac{1}{s}\sum^{s}_{i=1}\{Y[(\nu-1)s+i]-y_{\nu}(i)\}^{2},\label{eq:16}
\end{eqnarray}
for each segment ${\nu}$, ${\nu}$ = 1,…..,${N_{s}}$ and
\begin{eqnarray}
F^{2}(s,\nu)&\equiv&\frac{1}{s}\sum^{s}_{i=1}\{Y[N-(\nu-N_{s})s+i]-y_{\nu}(i)\}^{2}.\label{eq:17}
\end{eqnarray}
for each segment ${\nu}$ = ${N_{s}+ 1, N_{s}+ 2, …..2N_{s}}$.
  ${y_{\nu}(i)}$ is the least-square best fit line in segment
  ${\nu}$.\\
    The last step of the detrending process is average over all
segments of the time series to obtain the fluctuation function
that given as
\begin{eqnarray}
F(s)&equiv&\left[\frac{1}{2N_{s}}\sum^{2N_{s}}_{\nu=1}F^{2}(s,\nu)\right],\label{eq:18}
\end{eqnarray}
$F(s)$ will increase with increasing $s$, and it is only defined
for the segment length, ${s\geq4}$.  A log-log plot of $F(s)$
versus $s$ need to be to determine the scaling behaviors.
Therefore, the above steps are repeated several times to obtain a
set data of $F(s)$ versus s as shown in Fig.3. The slope of the
curve shows the scaling exponent ${\alpha}$, if the time series
are long-range power-law correlated. Hence, $F(s)$ and $s$ can be
related with a power-law relation which is given as
\begin{eqnarray}
F(s)&\sim&s^{\alpha}, \label{eq:19}
\end{eqnarray}
The scaling exponent can be deviated to a few category and is
summarized in Table III.
\begin{table}
\caption{\label{Tab:3} category of the scaling exponent,
${\alpha}$ with different processes.}
\begin{ruledtabular}
\begin{tabular}{|c|c|}
  Scaling exponent & Type od processes \\\hline
  ${0<\alpha<0.5}$ & Power-law anti-correlation \\
  ${\alpha=0.5}$ & White noise \\
  ${0.5<\alpha<1.0}$ & Long-range power-law correlation \\
  ${\alpha=1.0}$ & ${\frac{1}{f}}$ process \\
  ${\alpha=1.55}$ & Brownian motion \\ \hline
\end{tabular}
\end{ruledtabular}
\end{table}
\subsection*{RESULT(B)}
    To test the accuracy of the DFA algorithm which used in this work,
the algorithm is used to calculate the scaling exponent of three
known scaling exponent generated signals, which are Brownian
motion, persistence power-law, and anti-persistence power-law
process with Holder exponent of ${H = 0.50, H = 0.80}$, and ${H =
0.20}$ respectively. The obtained results are shown in Table IV.
The calculated DFA scaling exponents, ${\alpha}$ of DFA method are
consistent with the Holder exponent for the three generated
signals, and this verified the DFA algorithm is accurate to
produce the actual results. The result of graph $F(s)$ versus $s$
for three signals is shown in Fig. 4.\\
    The scaling exponent of Library of Congress's sending and
requesting time series are estimated with DFA method, and the
results are tabulated in Table V. The DFA method results show
these two signals exhibit cross over phenomenon at the segment
length of 60 and at 400 as shown in Fig. 5. It can be noticed that
scaling exponent ${\alpha}$ of these two signals are identical
with each others, which ${\alpha}$ change from white noise ${(s
\leq 60)}$ to ${\frac{1}{f}}$ process ${(s\leq 400)}$, and then,
to a process with${\alpha \approx 2.00}$, finally.

\begin{table}
\caption{\label{Tab:4} ${\alpha}$ of the persistence power-law
process, Brownian motion, and anti-persistence power-law process.}
\begin{ruledtabular}
\begin{tabular}{|c|c|c|}
  Time series & DFA Scaling Exponent, ${\alpha}$& ${\pm
  \alpha}$\\\hline
  Persistence Power-Law & 1.79 & 0.03 \\
  Brownian & 1.51 & 0.09 \\
  Anti-Persistence Power-Law & 1.17 & 0.10 \\ \hline
\end{tabular}
\end{ruledtabular}
\end{table}

\begin{table}
\caption{\label{Tab:5} ${\alpha}$ for the LOC(request) and
LOC(send).}
\begin{ruledtabular}
\begin{tabular}{|c|c|c|c|c|c|c|}
  Time series & ${\alpha_{1}}$ & ${\pm \alpha_{1}}$ & ${\alpha_{2}}$ & ${\pm \alpha_{2}}$ & ${\alpha_{3}}$ & ${\pm \alpha{3}}$
  \\\hline
  LOC(request) & 0.63 & 0.04 & 1.08 & 0.05 & 2.01 & 0.05\\
  LOC(send) & 0.65 & 0.03 & 1.18 & 0.05 & 1.95 & 0.02 \\ \hline
\end{tabular}
\end{ruledtabular}
\end{table}

\section{\label{V} TIME-SCALE ANALYSIS (TSA)}
    The previous described methods are based on linear log-log plot
which give only a single value of the \H, these methods are found
to be insufficient in estimating the locally time-varying Holder
exponent,$H(t)$.  The wavelet approach were a powerful tool to
solve this problem.  The wavelet transform (WT) is a tool which
can be function as a mathematical microscope that can well adapted
to reveal the hierarchy and governs the spatial distribution of
the singularities of multifractal measures.  We only consider the
continuous wavelet transform (CWT) in time-scale analysis in order
to estimate the $H(t)$. The CWT are defined as
\begin{eqnarray}
W_{x}(t,a;\psi)&=&\int^{+\infty}_{-\infty}x(s)\psi^{*}_{t,a}(s)ds,\label{eq:20}
\end{eqnarray}
where wavelet for different scale are defined as,
\begin{eqnarray}
\psi_{t,a}(s)&=&|a|^{\frac{1}{2}}\psi\left(\frac{s-t}{a}\right),\label{eq:21}
\end{eqnarray}
and $a$ is the scaling parameter and also ${a\propto
\frac{1}{\omega}}$. In this paper, we using Morlet wavelet in the
TSA and scalogram are defined as
\begin{eqnarray}
E_{x}&=&\int^{+\infty}_{-\infty}\int^{+\infty}_{-\infty}|W_{x}(t,a;\psi)|^{2}dt\frac{da}{a^{2}},\label{eq:22}
\end{eqnarray}
with $E_{x}$ is the energy of function $x$.  Therefore scalogram
is a energy distribution function of a signal or time series in
time-scale plane associated with $dt$$\frac{da}{a^{2}}$.
Considering a time series with uniform \H, which written as
\begin{eqnarray}
|x(s)-x(t)|&\leq& c|s-t|^{H},\label{eq:23}
\end{eqnarray}
where $c$ is a constant.  Applied CWT to $x(t)$ will form the
equation as,
\begin{eqnarray}
|W_{x}(t,a;\psi)|&\leq&c|a|^{H+\frac{1}{2}}\int^{+\infty}_{-\infty}|t|^{H}|\psi(t)|dt,\label{eq:24}
\end{eqnarray}
And the scalogram of this time series given by\cite {2}:
\begin{eqnarray}
\Omega_{scalo}(t,a)&\equiv&|W_{x}(t,a)|^{2}\sim|a|^{2H(t)+1};\label{eq:25}
\end{eqnarray}
when ${a \rightarrow 0}$.  From Eq. (25), one can estimate the
$H(t)$ , and also the global \H can be written as
\begin{eqnarray}
H_{global}&=&\frac{1}{T}\int^{T}_{0}H(t)dt, \label{eq:26}
\end{eqnarray}
Thus, TSA provide global \H and local $H(t)$.  Therefore TSA are
more powerful tool compare PSA and DFA in fractal analysis, since
most phenomena shown multifractal scaling behaviors.
\subsection*{RESULT(C)}
    The scalogram allow one to estimate the local $H(t)$ and global \H.
Fig. 6 show the graph of local $H(t)$ for LOC(request) and
LOC(send).  The red line represented the global \H for each time
series.  The result of the TSA for each time series are summarized
into table  VI.
\begin{table}
\caption{\label{Tab:6} Maximum, minimum value of $H(t)$, global \H
and \d  for LOC(request) and LOC(send).}
\begin{ruledtabular}
\begin{tabular}{|c|c|c|c|c|}
  Time series & $H(t)[min.]$ & $H(t)[max.]$ & Global \H & \d
  \\\hline
  LOC(request) & -0.49 & 1.48 & 0.32 & 1.68 \\
  LOC(send) & -0.26 & 1.15 & 0.27 & 1.73 \\ \hline
\end{tabular}
\end{ruledtabular}
\end{table}

\section{\label{VI} DISCUSSION}
    From the analysis results, proven that these two internet traffic
time series exhibit fractal characteristics with long-range
dependence. Therefore a previous increment of the time series will
affect the future increment,  or in other words both  internet
traffic time series behave like long-range memory phenomena, like
most in the nature.  Even though Fourier transform are using
harmonic basis function and have been shown the drawback for the
non-stationary signal processing, but PSA can be used for initial
measurements in fractal analysis for the nonstationary time series
like the objects we are studied.  From the PSA results, we get the
value for the \H  = 0.30+0.01 and 0.31+0.01 for LOC(request) and
LOC(send) repetitively.\\
    For the DFA results, show that LOC(request) and LOC(send) time
series exhibit crossover phenomenon within different segment
length $s$. This is probably due to the fact that on very short
times scale (starting time of requesting and sending files), the
internet traffic time series is dominated by highly uncorrelated
fluctuation process. As the time goes on, these signals exhibit
smoother fluctuation that reflect the intrinsic dynamic of many
electronic systems, which usually produce a $\alpha$ exponent
equal to one, and associate with the $\frac{1}{f}$ process like.\\
    Meanwhile TSA results show that these two internet traffic time
series are very complicated systems with local $H(t)$ cover from
negative value to positive value, which ${-0.49\leq H \leq 1.48}$
for LOC(request) and ${-0.26 \leq H \leq 1.15}$ for LOC(send).
Also seen that $H(t)$ for LOC(request) are more complex compare to
$H(t)$ for LOC(send). An explanation for the different complexity
of the $H(t)$ for both time series can be similar to the road
traffic at a gateway toward a metropolitans city.  For the
LOC(request), the data are coming from hundred of millions points
at the web network into a main gate at LOC server, this will
create an serious "traffic jam"  at the gateway of LOC server.
  Furthermore exist interaction between one incoming signal and
another incoming signal at the gateway during the period which the
incoming signal are overloaded, and caused the network congestion.
  As comparison, the LOC(send) are more regular because the data are
transfer from the main gateway to hundred of millions point at the
web network, this data transferring are more easy compare to the
incoming case.  Therefore the global \H value which are getting as
average value form $H(t)$ just an approximation, and give us the
coarse image for the time series dynamical behavior.  Since the
$H(t)$ for LOC(request) and LOC(send) are out of the range $( 0 <
H < 1)$, therefore these two internet traffic time series can be
threat as very complicated systems and encourage the further study
on its, and get a good quantitative description can advanced our
understanding of these two internet traffic time series.  However,
TSA provide us extra information compare to PSA and DFA, since it
give the local singularities multifractal behaviors, which allowed
us to study the detail behavior of the complex systems such like
the internet traffic time series.

\section{\label{VII} CONCLUSION}
    In this paper, we have examined the fractal characteristics and
long-range dependence in these two internet traffic time series.
We examined these LOC(request) and LOC(send) time series by three
techniques: power-spectral analysis(PSA), detrended fluctuation
analysis (DFA) and time-scale analysis(TSA).  Other techniques to
examined long-range dependence, not discussed in this paper,
include dispersional analysis\cite {11} and maximum-likelihood
estimator\cite{12}.  As summary, we find the following:-\\
(1) PSA quantify that (\b  : 1$<$\b$<$2), (\H : 0$<$\H$<$0.5), (\p
: -0.5$<$\p$<$0), and (\d : 1$<$\d$<$2).  PSA showed these two
internet traffic time series exhibit the fractal and long-range
dependence characteristics.\\
(2) We have used DFA method to analysis the networking signals,
and we find out that these signals exhibit crossover phenomenon at
the segment length of 60 and 400. Besides, signal of requesting
and sending have identical $\alpha$ exponent which show white
noise behavior for segment length of 60, $\frac{1}{f}$ process for
segment length of 400, and a smother process $(\alpha = 2.00)$ for
the entire signals.\\
(3) TSA quantify that ( Local $H(t)$ : -0.5$<$$H(t)$$<$ 1.5),
(Global \H : 0$<$\H$<$0.5) and (1$<$\d$<$2).  TSA showed that
LOC(request) and LOC(send) time series are two complicated time
series with local $H(t)$ out of the range in between 0 to 1.
Therefore these require advanced quantitative and qualitative
description of these signal to improve our understanding of the
internet traffic time series. In many ways, wavelets analysis are
the most effective method to perform the fractal analysis since it
can used for data sets that's are nonstationary and can perform
the multifractal measurements. According the analysis results, we
showed that the long-range dependence and fractal characteristics
exhibit in these LOC(request) and LOC(send) time series.  As the
value of H approach to zero, the systems became more complex.
Therefore we suggest that further fractal analysis and modeling
can be use in internet traffic time series to optimize the network
utilities.

\subsection*{ ACKNOWLEDGEMENT}
    The authors would like to than Sithi V.
Muniandy and Lim Swee Cheng for the long and thought-provoking
discussion in both the theoretical and practical application. K.B.
Chong would like to thank NUS and K.Y. Choo would like to thank
MMU for the partial financial support.

\begin{figure}
\caption{\label{Fig:1}The power-spectral
for(a)LOC(request),(b)LOC(send)}
\end{figure}

\end{document}